\documentclass[10pt, conference, compsocconf]{IEEEtran}
\IEEEoverridecommandlockouts

\newtheorem{lemma}{Lemma}

\newtheorem{theorem}[lemma]{Theorem}

\newtheorem{corollary}[lemma]{Corollary}

\usepackage{times}

\usepackage{cite}
\ifCLASSINFOpdf
\usepackage[pdftex]{graphicx}
\else
\fi
\usepackage[cmex10]{amsmath}
\usepackage{array}
\usepackage{mdwmath}
\usepackage{mdwtab}
\usepackage[tight,footnotesize]{subfigure}
\begin{document}
\title{Finding Associations and Computing Similarity via Biased Pair Sampling$^*$ \thanks{$^*$ This is an extended version of a paper that appeared at the IEEE International Conference on Data Mining, 2009. \copyright\  2009 IEEE. This version is {\bf superseded} by a full version that can be found at {\tt http://www.itu.dk/people/pagh/papers/mining-jour.pdf}, which contains stronger theoretical results and fixes a mistake in the reporting of experiments.}}


\author{\IEEEauthorblockN{Andrea Campagna and Rasmus Pagh}
\IEEEauthorblockA{IT University of Copenhagen, Denmark\\
Email: {\tt \{acam,pagh\}@itu.dk}}
%\and
% \IEEEauthorblockN{Authors Name/s per 2nd Affiliation (Author)}
% \IEEEauthorblockA{line 1 (of Affiliation): dept. name of organization\\
% line 2: name of organization, acronyms acceptable\\
% line 3: City, Country\\
% line 4: Email: name@xyz.com}
}


\maketitle

\begin{abstract}
Sampling-based methods have previously been proposed for the problem of finding interesting associations in data, even for low-support items. While these methods do not guarantee precise results, they can be vastly more efficient than approaches that rely on exact counting. However, for many similarity measures no such methods have been known. In this paper we show how a wide variety of measures can be supported by a simple {\em biased\/} sampling method. The method also extends to find high-confidence association rules. We demonstrate theoretically that our method is superior to exact methods when the threshold for ``interesting similarity/confidence'' is above the average pairwise similarity/confidence, and the average support is not too low. Our method is particularly good when transactions contain many items. We confirm in experiments on standard association mining benchmarks that this gives a significant speedup on real data sets (sometimes much larger than the theoretical guarantees). Reductions in computation time of over an order of magnitude, and significant savings in space, are observed.
\end{abstract}


\IEEEpeerreviewmaketitle

\section{Introduction}

A central task in data mining is finding associations in a binary relation. Typically, this is phrased in a ``market basket'' setup, where there is a sequence of baskets (from now on ``transactions''), each of which is a set of items. The goal is to find patterns such as ``customers who buy diapers are more likely to also buy beer''. There is no canonical way of defining whether an association is interesting --- indeed, this seems to depend on problem-specific factors not captured by the abstract formulation. As a result, a number of measures exist: In this paper we deal with some of the most common measures, including {\em Jaccard}~\cite{journals/tkde/CohenDFGIMUY01}, {\em lift}~\cite{Brin:1997:BMB,Aggarwal:1998:NFI}, {\em cosine}, and {\em all\_confidence}~\cite{conf/icdm/LeeKCH03,journals/tkde/Omiecinski03}. In addition, we are interested in high-confidence association rules, which are closely related to the {\em overlap coefficient\/} similarity measure. We refer to~\cite[Chapter 5]{datamining} for general background and discussion of similarity measures.

In the discussion we limit ourselves to the problem of binary associations, i.e., patterns involving pairs of items. There is a large literature considering the challenges of finding patterns involving larger item sets, taking into account the aspect of time, multiple-level rules, etc. While some of our results can be extended to cover larger item sets, we will for simplicity concentrate on the binary case. Previous methods rely on one of the following approaches: 

\begin{enumerate}
\item Identifying item pairs $(i,j)$ that ``occur frequently together'' in the transactions --- in particular, this means counting the number of co-occurrences of each such pair ---  or 
\item Computing a ``signature'' for each item such that the similarity of every pair of items can be estimated by (partially) comparing the item signatures.
\end{enumerate}

Our approach is different from both these approaches, and generally offers improved performance and/or flexibility.  In some sense we go directly to the desired result, which is the set of pairs of items with similarity measure above some user-defined threshold $\Delta$. Our method is {\em sampling\/} based, which means that the output may contain false positives, and there may be false negatives. However, these errors are rigorously understood, and can be reduced to any desired level, at some cost of efficiency --- our experimental results are for a false negative probability of less than 2\%. The method for doing sampling is the main novelty of this paper, and is radically different from previous approaches that involve sampling.

The main focus in many previous association mining papers has been on {\em space usage\/} and the number of {\em passes\/} over the data set, since these have been recognized as main bottlenecks. We believe that time has come to also carefully consider {\em CPU time}. A transaction with $b$ items contains $\binom{b}{2}$ item pairs, and if $b$ is not small the effort of considering all pairs is non-negligible compared to the cost of reading the item set. This is true in particular if data resides in RAM, or on a modern SSD that is able to deliver data at a rate of more than a gigabyte per second. One remedy that has been used (to reduce space, but also time) is to require {\em high support}, i.e., define ``occur frequently together'' such that most items can be thrown away initially, simply because they do not occur frequently enough (they are below the {\em support threshold\/}). However, as observed in~\cite{journals/tkde/CohenDFGIMUY01} this means that potentially interesting or useful associations (e.g.~correlations between genes and rare diseases) are not reported. In this paper we consider the problem of finding associations {\em without\/} support pruning. Of course, support pruning can still be used to reduce the size of the data set before our algorithms are applied.

In the following sections we first discuss the need for focusing on CPU time in data mining, and then elaborate on the relationship between our contribution and related works.

\medskip

\subsection{I/O versus CPU}

In recent years, the capacity of very fast storage devices has exploded. A typical desktop computer has 4--16 GB of RAM, that can be read (sequentially) at a speed of at least 800 million 32-bit words per second. The flash-based ioDrive Duo of Fusion-io offers up to over a terabyte of storage that can be read at around 400 million 32-bit words per second. 
Thus, even massive data sets can be read at speeds that make it challenging for CPUs to keep up. An 8-core system must, for example, process 100 million (or 50 million) items per core per second. At 3 GHz this is 33 clock cycles (or 66 clock cycles) per item. This means that any kind of processing that is not constant time per item (e.g., using time proportional to the size of the transaction containing the item) is likely to be CPU bound rather than I/O bound. For example, a hash table lookup requires on the order of 5-10 ns even if the hash table is L2 cache-resident (today less than 10 MB per core). This gives an upper limit of 100-200 million lookups per second in each core, meaning that any algorithm that does more than a dozen hash table operations per item (e.g. updating the count of some item pairs) is definitely CPU bound, rather than I/O bound.
In conclusion, we believe it is time to carefully consider optimizing internal computation time, rather than considering all computation as ``free'' by only counting I/Os or number of passes. Once CPU efficient algorithms are known, it is likely that the remaining bottleneck is I/O. Thus, we also consider I/O efficient versions of our algorithm.

% Intel Core 2 Duo 3 GHz, E8400 Wolfdale, L2 cache (6 MB) has 14 cycle latency, 4.7 ns.
% RAM 83 ns latency.

% Rasmus-Paghs-computer-2:~/papers/sse/src pagh$ icpc -msse3 -O3 linear.cpp -o linear            
% linear.cpp(23) : (col. 5) remark: LOOP WAS VECTORIZED.                                         
% Rasmus-Paghs-computer-2:~/papers/sse/src pagh$ ./linear Generating 2097152 pointer-value pairs of 2 x 4 bytes...                                                                             
% 100 linear probing sequences begin                                                             
% End: 209715100                                                                                 
% 0.250000 seconds                                                                               
%                                                                                                
% 100 double hashing sequences begin                                                             
% End: 419430200                                                                                 
% 2.030000 seconds                                                                               
 
% Note: Just over 100 million lookups per second.
% Note: No actual pointers in impl. Uses array of 2^21 4-byte integers = 8 MB. Size of L2 cache is 4 MB.

\subsection{Previous work}\label{sec:previous}

\subsubsection*{Exact counting of frequent item sets} 

The approach pioneered by the A-Priori algorithm~\cite{Agrawal+Mehta+Shafer+Srikant+Arning+Bollinger+1996a,VLDB94*487}, and refined by many others (see e.g.~\cite{conf/fimi/2003,conf/fimi/2004,Brin:1997:DIC,park.ea:effective-hash-based:95,Savasere:1995:EAM}), allows, as a special case, finding all item pairs $(i,j)$ that occur in more than $k$ transactions, for a specified threshold $k$. However, for the similarity measures we consider, the value of~$k$ must in general be chosen as a low constant, since even pairs of very infrequent items can have high similarity. This means that such methods degenerate to simply counting the number of occurrences of all pairs, spending time $\Theta(b^2)$ on a transaction with $b$ items. Also, generally the space usage of such methods (at least those requiring a constant number of passes over the data) is at least 1 bit of space for each pair that occurs in {\em some\/} transaction.

The problem of counting the number of co-occurrences of all item pairs is in fact equivalent to the problem of multiplying sparse 0-1 matrices. To see this, consider the $n \times m$ matrix A in which each row A$_i$ is the incidence vector
having $1$ in position $p$ iff the $i$th element in the set of items appears in the $p$th transaction. Each entry $\tilde{\textsc{A}}_{i,j}$ of the $n \times n$ matrix $\tilde{\textsc{A}}=
\textsc{A}\times\textsc{A}^T$ represents the number of transactions in which the pair $(i,j)$ appears. The best theoretical algorithms for (sparse) matrix multiplication~\cite{DBLP:conf/icdt/AmossenP09,journals/jsc/CoppersmithW90,mm_yuster_zwick} scale better than the A-Priori family of methods as the transaction size gets larger, but because of huge constant factors this is so far only of theoretical interest.

\subsubsection*{Sampling transactions} 

Toivonen~\cite{Toivonen:1996:SLD} investigated the use of sampling to find candidate frequent pairs $(i,j)$: Take a small, random subset of the transactions and see what pairs are frequent in the subset. This can considerably reduce the memory used to actually count the number of occurrences (in the full set), at the cost of some probability of missing a frequent pair. This approach is good for high-support items, but low-support associations are likely to be missed, since few transactions contain the relevant items.

\subsubsection*{Locality-sensitive hashing} 

Cohen et al.~\cite{journals/tkde/CohenDFGIMUY01} proposed the use of another sampling technique, called {\em min-wise independent hashing\/}, where a small number of occurrences of each item (a ``signature'') is sampled. This means that occurrences of items with low support are more likely to be sampled. As a result, pairs of (possibly low-support) items with high {\em jaccard coefficient\/} are found --- with a probability of false positives and negatives. A main result of~\cite{journals/tkde/CohenDFGIMUY01} is that the time complexity of their algorithm is proportional to the sum of all pairwise jaccard coefficients, plus the cost of initially reading the data. Our main result has basically the same form, but has the advantage of supporting a wide class of similarity measures.

Min-wise independent hashing belongs to the class of {\em locality-sensitive\/} hashing methods~\cite{STOC::IndykMRV1997}. Another such method was described by Charikar~\cite{charikar}, who showed how to compute succinct signatures whose Hamming distance reflects angles between incidence vectors. This leads to an algorithm for finding item pairs with cosine similarity above a given threshold (again, with a probability of false positives and negatives), that uses linear time to compute the signatures, and $\Theta(n^2)$ time to find the similar pairs, where $n$ is the number of distinct items in all transactions. Charikar also shows that many similarity measures, including some measures supported by our algorithm, cannot be handled using the approach of locality-sensitive hashing.

\subsubsection*{Deterministic signature methods} 

In the database community, finding all pairs with similarity above a given threshold is sometimes referred to as a ``similarity join.'' Recent results on similarity joins include~\cite{conf/vldb/ArasuGK06,conf/icde/ChaudhuriGK06,conf/icde/XiaoWLS09,conf/www/XiaoWLY08}. While not always described in this way, these methods can be seen as deterministic analogues of the locality-sensitive hashing methods, offering {\em exact\/} results. The idea is to avoid computing the similarity of every pair by employing succinct ``signatures'' that may serve as witnesses for low similarity. Most of these methods require the signatures of every pair of items to be (partially) compared, which takes $\Omega(n^2)$ time. However, the worst-case asymptotic performance appears to be no better than the A-Priori family of methods. A similarity join algorithm that runs faster than $\Omega(n^2)$ in some cases is described in~\cite{conf/vldb/ArasuGK06}. However, this algorithm exhibits a polynomial dependence on the maximum number $k$ of differences between two incidence vectors that are considered similar, and for many similarity measures the relevant value of $k$ may be linear in the number $m$ of transactions.

%%%%%%%%%%%%%%%%%%%%%%%%%%%%%%%%%%%%%%%%%%%%%%%%%%%%%

\subsection{Our results}

In this paper we present a novel sampling technique to handle a variety of measures (including jaccard, lift, cosine, and all\_confidence), even finding similar pairs among low support items. The idea is to sample a subset of all pairs $(i,j)$ occurring in the transactions, letting the sampling probability be a function of the supports of $i$ and $j$. For a parameter~$\mu$, the probability is chosen such that each pair with similarity above a threshold $\Delta$ (an ``interesting pair'') will be sampled at least $\mu$ times, in expectation, while we do not expect to see a pair $(i,j)$ whose measure is significantly below $\Delta$. A na\"ive implementation of this idea would still use quadratic time for each transaction, but we show how to do the sampling in {\em near-linear time\/} (in the size of the transaction and number of sampled pairs).

The number of times a pair is sampled follows a binomial distribution, which allows us to use the sample to infer which pairs are likely to have similarity above the threshold, with rigorous bounds on false negative and false positive probabilities. We show that the time used by our algorithm is (nearly) linear in the input size and in the the sum of all pairwise similarities between items, divided by the threshold $\Delta$. This is (close to) the best complexity one could hope for with no conditions on the distribution of pairwise similarities. Under reasonable assumptions, e.g.~that the average support is not too low, this gives a speedup of a factor $\Omega(b/\log b)$, where~$b$ is the average size of a transaction.

We show in extensive experiments on standard data sets for testing data mining algorithms that our approach (with a $1.8\%$ false negative probability) gives speedup factors in the vicinity of an order of magnitude, as well as significant savings in the amount of space required, compared to exact counting methods.
We also present evidence that for data sets with many distinct items, our algorithm may perform significantly less work than methods based on locality-sensitive hashing.

\subsection{Notation}\label{sec:notation}

Let $T_1,\dots,T_m$ be a sequence of transactions, $T_j\subseteq [n]$. For $i=1,\dots,n$ let $S_i = \{ j \; | \; i\in T_j\}$, i.e., $S_i$ is the set of occurrences of item $i$.

We are interested in finding associations among items, and consider a framework that captures the most common measures from the data mining literature.
Specifically, we can handle a similarity measure $s(i,j)$ if there exists a function $f: {\bf N}\times {\bf N} \times {\bf R}_+ \rightarrow {\bf R}_+$  that is non-increasing in all parameters, and such that:
$$ |S_i \cap S_j| \,f(|S_i|,|S_j|,s(i,j)) = 1 \enspace . $$

In other words, the similarity should be the solution to an equation of the form given above. Fig.~\ref{table:functions} shows particular measures that are special cases. 
The monotonicity requirements on $f$ hold for any reasonable similarity measure: increasing $|S_i \cap S_j|$ should not decrease the similarity, and adding an occurrence of $i$ or $j$ should not increase the similarity unless $|S_i \cap S_j|$ increases.
In the following we assume that $f$ is computable in constant time, which is clearly a reasonable assumption for the measures of Fig.~\ref{table:functions}.

\begin{figure}
\begin{center}
\begin{tabular}{c|c|c|}
{\bf Measure} & $s(i,j)$ & $f(|S_i|,|S_j|,s)$\\
\hline
\hline
{\LARGE\phantom{(}}lift & $\frac{|S_i\cap S_j|}{|S_i| |S_j|}$ & $s^{-1} m/(|S_i|\cdot|S_j|)$\\
\hline
{\LARGE\phantom{(}}cosine & $\frac{|S_i\cap S_j|}{\sqrt{|S_i| |S_j|}}$ & $s^{-1} /\sqrt{|S_i|\cdot |S_j|}$\\
\hline
{\LARGE\phantom{(}}jaccard & $\frac{|S_i\cap S_j|}{|S_i \cup S_j|}$ & $\frac{1+s}{s} /(|S_i| + |S_j|)$\\
\hline
{\LARGE\phantom{(}}all\_confidence & $\frac{|S_i\cap S_j|}{\max(|S_i|,|S_j|)}$ & $s^{-1} /\max(|S_i|,|S_j|)$\\
\hline
{\LARGE\phantom{(}}dice & $\frac{|S_i\cap S_j|}{|S_i|+|S_j|}$ & $s^{-1} /(|S_i|+|S_j|)$\\
\hline
{\LARGE\phantom{(}}overlap\_coef & $\frac{|S_i\cap S_j|}{\min(|S_i|,|S_j|)}$ & $s^{-1} /\min(|S_1|,|S_2|)$\\
%\hline
% confidence($i\Rightarrow j$) & $s^{-1} /|S_i|$\\
\hline
\end{tabular}
\caption{\normalsize\em Some measures covered by our algorithm and the corresponding functions. Note that finding all pairs with overlap coefficient at least $\Delta$ implies finding all association rules with confidence at least $\Delta$.
%We note that the last measure, confidence of an association rule, is asymmetric.
}\label{table:functions}
\end{center}
\end{figure}

%\newpage
\section{Our algorithm}

The goal is to identify pairs $(i,j)$ where $s(i,j)$ is ``large''. Given a user-defined threshold~$\Delta$ we want to report the pairs where $s(i,j)\geq \Delta$. We observe that all measures in Fig.~\ref{table:functions} are symmetric, so it suffices to find all pairs $(i,j)$ where $|S_i| \leq |S_j|$, $i\ne j$, and $s(i,j)\geq \Delta$.

\subsection{Algorithm idea} Our algorithm is randomized and finds each qualifying pair with probability~$1-\varepsilon$, where $\varepsilon > 0$ is a user-defined error probability. The algorithm may also return some false positives, but each false positive pair is likely to have similarity within a small constant factor of $\Delta$. If desired, the false positives can be reduced or eliminated in a second phase, but we do not consider this step here. 

The basic idea is to randomly sample pairs of items that occur together in some transaction such that for any pair $(i,j)$ the expected number times it is sampled is a strictly increasing function of $s(i,j)$. Indeed, in all cases except the jaccard measure it is simply proportional to $s(i,j)$. We scale the sampling probability such that for all pairs with $s(i,j)\geq\Delta$ we expect to see at least $\mu$ occurrences of $(i,j)$, where $\mu$ is a parameter (defined later) that determines the error probability.

\begin{figure}[t]
\begin{tabbing}
xx\=xx\=xx\=xx\=xx\=\kill
{\bf procedure} {\sc BiSam}$(T_1,\dots,T_m;f,\mu,\Delta)$\+\\
$c:=${\sc ItemCount}$(T_1,\dots,T_m)$;\\
$M:=\emptyset$;\\
{\bf for} $t:=1$ to $m$ {\bf do}\+\\
sort $T_t[]$ s.t.~$c(T_t[j])\leq c(T_t[j+1])$ for $1\leq j < |T_t|$;\\
let $r$ be a random number in $[0;1)$;\\
{\bf for} $i:=1$ to $|T_t|$ {\bf do}\+\\
j:=i+1;\\
{\bf while} $j\leq |T_t|$ and $f(c(T_t[i]),c(T_t[j]),\Delta)\mu > r$ {\do}\+\\
$M:=M\cup \{(T_t[i],T_t[j])\}$;\\
j:=j+1;\-\\
{\bf end}\-\\
{\bf end}\-\\
{\bf end}\\
$R=\emptyset$;\\
{\bf for} $(i,j)\in M$ {\bf do}\+\\
{\bf if} $M(i,j) > \mu/2$ {\bf or} $M(i,j)f(c(i),c(j),\Delta)\geq 1$ {\bf then}\+\\
 $R:=R\cup \{(i,j)\}$;\-\-\\
%{\bf end}\\
{\bf return} $R$;\-\\
{\bf end}\\
\end{tabbing}
\vspace{-5mm}
\caption{\normalsize\em Pseudocode for the {\sc BiSam } algorithm. The procedure {\sc ItemCount}$(\cdot)$ returns a function (hash map) that contains the number of occurrences of each item. $T_t[j]$ denotes the $j$th item in transaction~$t$. $M$ is a {\em multi\/}set that is updated by inserting certain randomly chosen pairs $(i,j)$. The number of occurrences of a pair $(i,j)$ is denoted $M(i,j)$.\label{fig:pseudocode}}
\end{figure}

% http://de.wikipedia.org/wiki/Bisamratte
% English: Muskrat
\subsection{Implementation} Fig.~\ref{fig:pseudocode} shows our algorithm, called {\sc BiSam} (for {\em bi\/}ased {\em sam\/}pling). 
The algorithm iterates through the transactions, and for each transaction $T_t$ adds a subset of $T_t\times T_t$ to a multiset $M$ in time that is linear in $|T_t|$ and the number of pairs added. We use $T_t[i]$ to denote the $i$th item in $T_t$. Because $f$ is non-increasing and $T_t$ is sorted according to the order induced by $c(\cdot)$ we will add $(T_t[i],T_t[j])\in T_t\times T_t$ if and only if $f(c(T_t[i]),c(T_t[j]),\Delta) \mu > r$. The second loop of the algorithm builds an output set containing those pairs $(i,j)$ that either occur at least $\mu/2$ times in $M$, or where the number of occurrences in $M$ imply that $s(i,j)\geq\Delta$ (with probability~$1$).

The best implementation of the subprocedure {\sc ItemCount} depends on the relationship between available memory and the number $n$ of distinct items. If there is sufficient internal memory, it can be efficiently implemented using a hash table. For larger instances, a sort-and-count approach can be used (Section~\ref{sec:IO}).
The multiset $M$ can be represented using a hash table with counters (if it fits in internal memory), or more generally by an external memory data structure. In the following we first consider the standard model (often referred to as the ``RAM model''), where the hash tables fit in internal memory, and assume that each insertion takes constant time. Then we consider the I/O model, for which an I/O efficient implementation is discussed. As we will show in Section~\ref{sec:errorprob}, a sufficiently large value of $\mu$ is 
 $8 \ln(1/\varepsilon)$. Fig.~\ref{table:errorprob} shows more exact, concrete values of $\mu$ and corresponding false positive probabilities $\varepsilon$.

\bigskip

{\bf Example.} Suppose {\sc ItemCount} has been run and the supports of items 1--6 are as shown in Fig.~\ref{fig:example}.
\begin{figure}
 \begin{center}
  \begin{tabular}{c|c||c|c|}
   item & occurences & item & occurrences \\
   \hline
   $i$ & $c(i)$ & $i$ & $c(i)$  \\
   \hline
   \hline
   1 & 60 & 4 & 5 \\
   2 & 60 & 5 & 5 \\
   3 & 50 & 6 & 3 \\
  \hline
  \end{tabular}
 \end{center}
\caption{\normalsize\em Items in the example, with corresponding {\sc ItemCount} values.}\label{fig:example}
\end{figure}
Suppose now that the transaction $T_t = \{6,5,4,3,2,1\}$ is given. Note that its items are written according to the number of occurrences of each item. Assuming the similarity measure is {\it cosine}, $\mu=10$, $\Delta=0.7$, and $r$ for this transaction equal to 0.9, our algorithm would select from $T_t\times T_t$ the pairs shown in Fig.~\ref{table:steptwo}.
 \begin{figure}
 \begin{center}
  \begin{tabular}{c|c|c||c|c|c|}
   $i$ & $j$ & $f(c(i),c(j),\Delta)$ & $i$ & $j$ & $f(c(i),c(j),\Delta)$\\
   \hline
   \hline
   6 & 5 & 0.37 & 6 & 2 & 0.11 \\
   6 & 4 & 0.37 &  6 & 1 & 0.11 \\
   6 & 3 & 0.12 & 5 & 4 & 0.28\\
  \hline
  \end{tabular}
 \end{center}
\caption{\normalsize\em Pairs selected from $T_t$ in the example. Notice that after realizing the pair $(5,3)$ does not satisfy the inequality $f(c(5),c(3),\Delta)\mu > r$, the algorithm will not take into account the pairs $(5,2)$ and $(5,1)$.}\label{table:steptwo}
\end{figure}

Suppose that after processing all transactions the pair $(6,5)$ occurs 3 times in $M$, $(6,4)$ occurs twice in $M$, $(6,1)$ occurs once in $M$, and $(5,4)$ occurs 4 times in $M$. Then the algorithm would output the pairs: $(6,5)$ (since $M(6,5)<\mu/2$ but $M(6,5) f(3,5,0.7) > 1$), and $(5,4)$ (same situation as before).

\section{Analysis of running time}\label{sec:analysis}

Our main lemma is the following:

\medskip

\begin{lemma}\label{lem:main}
For all pairs $(i,j)$, where $i\ne j$ and $c(i)\leq c(j)$, if $f(c(i),c(j),\Delta)\mu < 1$ then at the end of the procedure, $M(i,j)$ has binomial distribution with $|S_i\cap S_j|$ trials and mean $$|S_i\cap S_j| f(|S_i|,|S_j|,\Delta) \mu .$$ If $f(c(i),c(j),\Delta)\mu \geq 1$ then at the end of the procedure $M(i,j) = |S_i\cap S_j|$.
\end{lemma}
\medskip
\begin{proof}
As observed above, the algorithm adds the pair $(i,j)$ to $M$ in iteration $t$ if and only if $(i,j)\in T_t\times T_t$ and $f(c(i),c(j),\Delta)\mu > r$, where $r$ is the random number in $[0;1)$ chosen in iteration $t$. This means that for every 
$t\in S_i\cap S_j$ we add $(i,j)$ to $M$ with probability $\min(1,f(c(i),c(j),\Delta)\mu)$. In particular, $M(i,j) = |S_i\cap S_j|$ for $f(c(i),c(j),\Delta) \mu \geq 1$. Otherwise, since the value of $r$ is independently chosen for each $t$, the distribution of $M(i,j)$ is binomial with $|S_i\cap S_j|$ trials and mean $|S_i\cap S_j| f(c(i),c(j),\Delta) \mu$.
\end{proof}

\medskip

Looking at Fig.~\ref{table:functions} we notice that for the jaccard similarity measure $s(i,j) = \frac{|S_i\cap S_j|}{|S_i]+|S_j|-|S_i\cap S_j|}$, the mean of the distribution is 
$$\frac{|S_i\cap S_j|}{|S_i]+|S_j|}\frac{1+\Delta}{\Delta}\mu = \mu \frac {s(i,j) (1+\Delta)}{(1+s(i,j)) \Delta} \leq 2 \mu\, s(i,j) / \Delta,$$
where the inequality uses $s(i,j),\Delta \in [0;1]$. For all other similarity measures the mean of the binomial distribution is $\mu\, s(i,j)/\Delta$. As a consequence, for all these measures, pairs with similarity below $(1-\varepsilon) \Delta$ will be counted exactly, or sampled with mean $(1-\Omega(\varepsilon)) \mu$. Also notice that for all the measures we consider, 
$$|S_i\cap S_j| f(|S_i|,|S_j|,\Delta) = O(s(i,j)/\Delta) .$$

We provide a running time analysis both in the standard (RAM) model and in the I/O model of Aggarwal and Vitter~\cite{MR90k:68029}. In the latter case we present an external memory efficient implementation of the algorithm, {\sc IOBiSam}. Let $b$ denote the average number of items in a transaction, i.e., there are $bm$ items in total. Also, let $z$ denote the number of pairs reported by the algorithm.

\subsection{Running time in the standard model}\label{sec:RAM}

The first and last part of the algorithm clearly runs in expected time $O(mb+z)$. The time for reporting the result is dominated by the time used for the main loop, but analyzing the complexity of the main loop requires some thought. The sorting of a transaction with~$b_1$ items takes $O(b_1\log b_1)$ time, and in particular the total cost of all sorting steps is $O(mb\log n)$.\footnote{We remark that if $O(mb)$ internal memory is available, two applications of radix sorting could be used to show a theoretically stronger result, by sorting all transactions in $O(mb)$ time, following the same approach as the external memory variant.}

What remains is to account for the time spent in the {\bf while} loop. We assume that $|S_i\cap S_j| f(|S_i|,|S_j|,\Delta) = O(s(i,j)/\Delta)$, which is true for all the measures we consider. The time spent in the while loop is proportional to the number of items sampled, and according to Lemma~\ref{lem:main} the pair $(i,j)$ will be sampled $|S_i\cap S_j| f(|S_i|,|S_j|,\Delta) \mu = O(\mu\, s(i,j)/\Delta)$ times in expectation if $f(c(i),c(j),\Delta)\mu < 1$, and $|S_i\cap S_j|$ times otherwise. In both cases, the expected number of samples is $O(s(i,j)\frac {\mu}{\Delta})$. Summing over all pairs we get the total time complexity.

\medskip 

\begin{theorem}\label{teo:RAM}
Suppose we are given transactions $T_1,\dots,T_m$, each a subset of $[n]$, with $mb$ items in total,
and that $f$ is the function corresponding to the similarity measure $s$. Also assume that
$$|S_i\cap S_j| f(|S_i|,|S_j|,\Delta) = O(s(i,j)/\Delta).$$
Then the expected time complexity of {\sc BiSam}$(T_1,\dots,T_m;f,\mu,\Delta)$ in the standard model is:
\begin{equation}\label{eq:RAMA}
  O\left(mb\log n + \frac{\mu}{\Delta}\sum_{1\leq i<j\leq n} s(i,j)\right) \enspace .
\end{equation}
\end{theorem}

\medskip

\paragraph{Discussion}
This result is close to the best we could hope for with no condition on the distribution of pairwise similarities. The first term is near-linear in the input size, and the output size $z$ may be as large as $\Omega(\Delta^{-1} \sum_{1\leq i<j\leq n} s(i,j))$. This happens if the average similarity among the pairs reported is $O(\Delta)$, and the total similarity among other pairs is low and does not dominate the sum. For such inputs, the algorithm runs in $O(mb \log n + \mu z)$ time, and clearly $\Omega(mb + z)$ time is needed by any algorithm.

A comparison can be made with the complexity of schemes counting the occurrences of all pairs. Such methods use time $\Omega(mb^2)$, which is a factor $b/\log n$ larger than the first term. In fact, the difference will be larger if the distribution of transaction sizes is not even --- and in particular the difference in time will be at least a factor $b/\log b$ (but this requires a more thorough analysis).
 Since ususally one is interested in reporting the highly similar pairs, the condition that $\Delta$ is greater than the average similarity $\sum_{1\leq i<j\leq n} s(i,j) / \binom{n}{2}$ is frequently true. (In fact, one could imagine that $\Delta$ would in many cases be much greater than the average similarity.) From the above we can obtain the following simple (in some cases pessimistic) upper bound on the time complexity:

\medskip

\begin{corollary}
 If $\Delta$ is not chosen smaller than the average pairwise similarity, the expected time complexity of {\sc BiSam} is $O(m b \log n + \mu n^2)$.
\end{corollary}

\medskip

This means that under the assumption of the corollary we win a factor of at least $\min(b/\log b,\tfrac{m}{\mu}(\tfrac{b}{n})^2)$ compared to the exact counting approach. Note in this context that $\mu$ can be chosen as a small value (e.g., $\mu = 15$ in our experiments). In most of our experiments the first of the two terms (the counting phase) dominated the time complexity. However, we also found that for some data sets with mainly low-support items, the second term dominated.
If we let $\sigma = mb/n$ denote the average support, the speedup can be expressed as $\Omega(b \min(1/\log b,\tfrac{\sigma}{\mu n}))$. That is, the second term dominates if the average support is below roughly $\mu n / \log b$.

\paragraph{Independent items}

As further evidence for (or explanation of) why the time complexity of the second term may be close to linear, we consider an input where each item $i$ appears in a given transaction with probability $p_i$, independently of all other items. Thus, the probability that distinct items $i$ and $j$ appear in a transaction is $p_i p_j$. 
We observe that each similarity measure $s(i,j)$ in Fig.~\ref{table:functions}, with the exception of {\em lift}, satisfies $s(i,j)\leq \bar{s}(i,j)$, where $\bar{s}(i,j)=\frac{|S_i\cap S_j|}{|S_i|}+\frac{|S_i\cap S_j|}{|S_j|}$. Thus, we get an upper bound on running time for these measures by considering the similarity measure $\bar{s}(i,j)$. Observe that the expected value of $\bar{s}(i,j)$ is $p_i+p_j$ by linearity of expectation.
Hence, the expected sum of similarities is:
$$\sum_{i=1}^n \sum_{j=i+1}^n p_i + p_j \leq \sum_{i=1}^n p_i n + \sum_{j=1}^n n p_j  = 2n \enspace .$$
This means that the running time of {\sc BiSam} is indeed $O(mb\log n + n/\Delta)$ for independent items.

\subsection{Running time in the I/O model}\label{sec:IO}

We now present {\sc IOBiSam}, an I/O efficient implementation of the {\sc BiSam} algorithm. The rest of the paper can be read independently of this section. As before, we assume that the similarity measure is such that $|S_i\cap S_j| f(|S_i|,|S_j|,\Delta) = O(s(i,j)/\Delta)$

In order to compute the support of each item, which means computing the {\sc ItemCount} function, a sorting of the dataset's items is carried out. It is necessary to keep track of which transaction each item belongs to. To compute the sorted list of items, O$(\frac{N}{B}\log_{\frac{M}{B}}\frac{N}{M})$ I/Os are needed~\cite{MR90k:68029}, where $N = mb$ is the number of pairs $c=\langle$item, Transaction ID$\rangle$, $M$ is the number of such pairs that fit in memory, and $B$ is the number of pairs that fit in a memory page. When the items are sorted, it is trivial to compute the number of occurrences of each item, so it takes just $O(\frac{N}{B})$ I/Os to compute and store the tuples $c\langle$item,support,Transaction ID$\rangle$. In the following, let $\tilde{C}$ be the set of such tuples written to disk.

We then sort the tuples according to transaction ID, and secondarily according to support, again using $O(\frac{N}{B}\log_{\frac{M}{B}}\frac{N}{M})$ I/Os. This gives us each transaction in sorted order, according to item supports. Assuming that each transaction fits in main memory\footnote{The assumption is made only for simplicity of exposition, since the result holds also without this assumption.} it is simple to determine which pairs satisfy the inequality $f(c(T_t[i]),c(T_t[j]),\Delta)\mu > r$. When a pair satisfies the inequality, it is buffered in an output page in memory, together with the item supports. Once the page is filled, it is flushed to external memory. The total cost of this phase is $O(\frac{N+N'}{B})$ I/Os for the flushings and reads, where $N^{'}$ is the total number of pairs satisfying the inequality (i.e., the number of samples taken). As before, the expected value of $N'$ is $O(\frac{\mu}{\Delta} \sum_{1 \leq i < j \le n} s(i,j))$. Finally, we spend $O(\frac{N'}{B}\log_{\frac{M}{B}}\frac{N'}{M})$ I/Os to sort the sampled pairs (according to e.g.~lexicographical order). Then it is easy to compute $M(i,j)$, i.e., the number of times each pair $(i,j)$ has been sampled by the algorithm, using $O(\frac{N^{'}}{B})$ I/Os. The final step is to output all the pairs satisfying the condition:
$$M(i,j) > \mu/2 \text{\bf\ or } M(i,j)f(c(i),c(j),\Delta)\geq 1,$$
which needs $O(\frac{N'}{B})$ I/Os. We observe that this cost is dominated by the cost of previous operations.
The most expensive steps are the sorting steps, whose total input has size $O(N+N')$, implying that the following theorem holds:

\medskip

\begin{theorem}
 Suppose we are given transactions $T_1,\dots,T_m$, each a subset of $[n]$, with $N = mb$ items in total,
and $f$ is the function corresponding to the similarity measure $s$. Also assume $|S_i\cap S_j| f(|S_i|,|S_j|,\Delta) = O(s(i,j)/\Delta)$. For $N'= O(\frac{\mu}{\Delta}{\sum_{1 \leq i < j \le n} s(i,j)})$, the expected complexity of {\sc IOBiSam}$(T_1,\dots{},T_m;f,\mu,\Delta)$ in the I/O model is
 $$ O\left( \tfrac{N+N'}{B}\log_{\frac{M}{B}}\left(\tfrac{N+N'}{M}\right) \right) \text{ I/Os } \enspace . $$
\end{theorem}\label{teo:IOmodel}

\section{Analysis of error probability}\label{sec:errorprob}

{\bf False negatives.} We first bound the probability that a pair $(i,j)$ with $s(i,j)\geq \Delta$ is not reported by the algorithm. This happens if $M(i,j)\leq \mu/2$ and $M(i,j) f(c(i),c(j),\Delta) < 1$. If $f(c(i),c(j),\Delta)\mu \geq 1$ then the pair $(i,j)$ is reported with probability~$1$. Otherwise, since $M(i,j)$ has binomial distribution, it follows from Chernoff bounds (see e.g.~\cite[Theorem 4.2]{MR96i:65003} with $\delta = 1/2$) that the probability of the former event is at most $\exp(-\delta^2 \mu / 2) = \exp(-\mu/8)$. Solving for $\mu$ this means that we have error probability at most $\varepsilon$ if $\mu \geq 8 \ln(1/\varepsilon)$.  This bound is pessimistic, especially when $\varepsilon$ is not very small. Tighter bounds can be obtained using the Poisson approximation to the binomial distribution, which is known to be precise when the number of trials is not too small (e.g., at least 100). Fig.~\ref{table:errorprob} shows some values of $\mu$ and corresponding false negative probabilities, using the Poisson approximation.

\begin{figure}
\begin{center}
\begin{tabular}{c|l|l|}
{\bf $\mu$} & {\bf $\varepsilon$}  & {\bf $\varepsilon'$}\\
\hline
\hline
3 & 0.199 & 0.0498\\
5 & 0.125 & 0.00674\\
10 & 0.0671 & 0.0000454\\
15 & 0.0180 & $< 10^{-6}$\\
20 & 0.0108 & $< 10^{-8}$\\
30 & 0.00195 & $< 10^{-13}$\\
\hline
\end{tabular}
\caption{\normalsize\em Values of $\mu$ and corresponding error probabilities $\varepsilon$. The error probabilities $\varepsilon'$ are for the variant of the algorithm where we return the whole multiset $M$, and use a different method to filter false positives (see Section~\ref{sec:variants}).}\label{table:errorprob}
%Source: http://stattrek.com/Tables/Poisson.aspx
\end{center}
\end{figure}

\medskip

\begin{figure}
\begin{center}
\includegraphics[width=.9\linewidth]{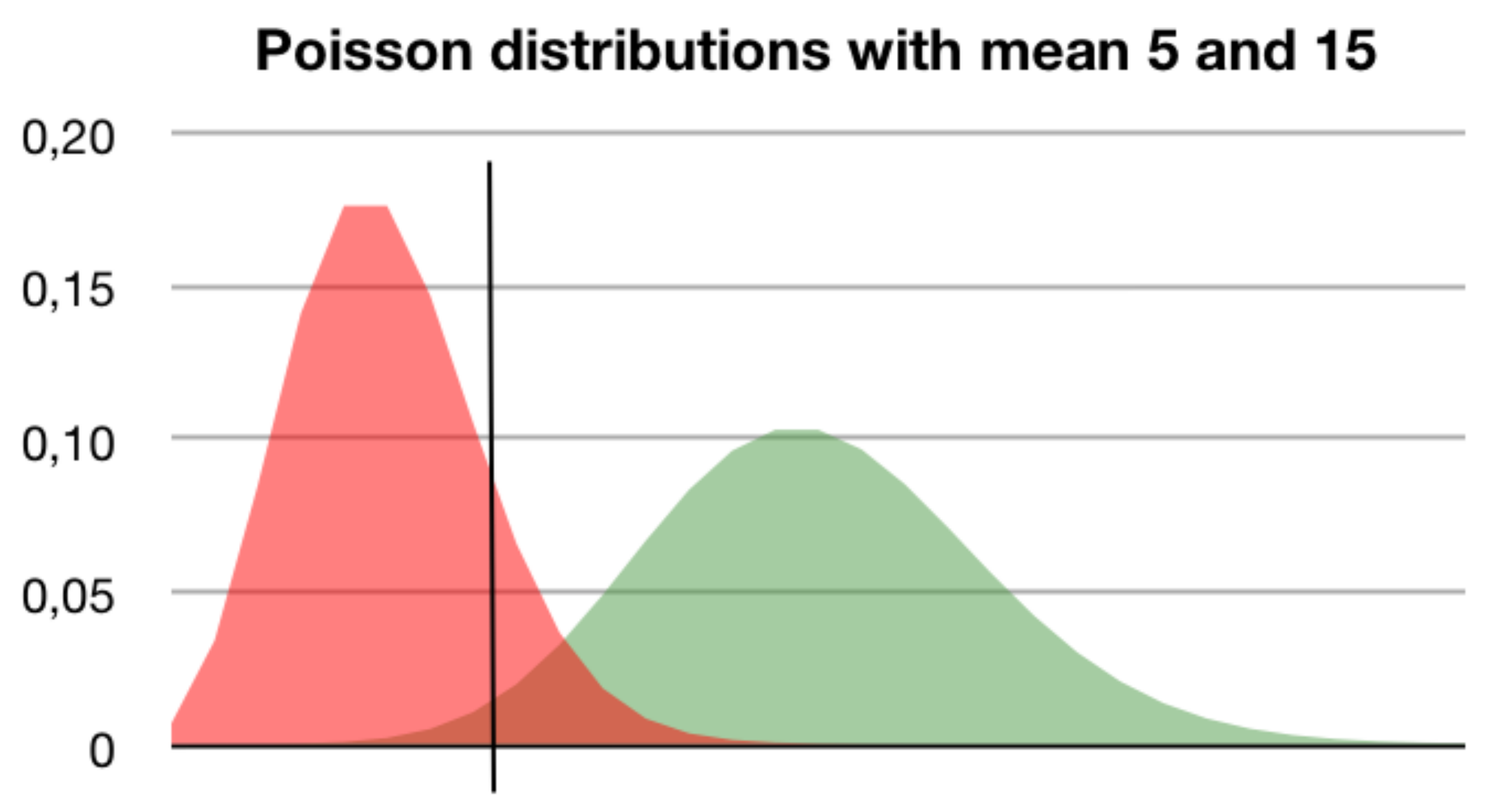}
\caption{\normalsize\em Illustration of false negatives and false positives for $\mu=15$. The leftmost peak shows the probability distribution for the number of samples of a pair $(i,j)$ with $s(i,j)=\Delta/3$. With a probability of around 13\% the number of samples is above the threshold (vertical line), which leads to the pair being reported (false positive). The rightmost peak shows the probability distribution for the number of samples of a pair $(i,j)$ with $s(i,j)=\Delta$. The probability that this is below the threshold, and hence not reported (false negative), is around 1.8\%.}\label{fig:poisson}
% Poisson.numbers
\end{center}
\end{figure}

%\noindent
{\bf False positives.} The probability that a pair $(i,j)$ with $s(i,j) < \Delta$ is reported depends on how far the mean $|S_i\cap S_j| f(|S_i|,|S_j|,\Delta) \mu$ is from $\mu$. With the exception of the jaccard measure, all measures we consider have mean $\mu\, s(i,j) / \Delta$. In the following we assume this is the case (a slightly more involved analysis can be made for the jaccard measure). If the ratio $s(i,j)/\Delta$ is close to $1$, there is a high probability that the pair will be reported. However, this is not so bad since $s(i,j)$ is close to the threshold $\Delta$. On the other hand, when $s(i,j) / \Delta$ is close to zero we would like the probability that $(i,j)$ is reported to be small. Again, we may use the fact that either $f(c(i),c(j),\Delta)\mu \geq 1$ (in which case the pair is exactly counted and reported with probability~$0$), or $M(i,j)$ has binomial distribution with mean $s(i,j)\frac {\mu}{\Delta}$. For $s(i,j)<\Delta/2$ we can use Chernoff bounds, or the Poisson approximation, to bound the probability that $M(i,j)>\mu/2$. Fig.~\ref{fig:poisson} illustrates two Poisson distributions (one corresponding to an item pair with measure three times below the threshold, and one corresponding to an item pair with measure at the threshold).

Actually, the number $\mu / 2$ in the reporting loop of the {\sc BiSam} algorithm is just one possible choice in a range of possible trade-offs between the number of false positives and false negatives. As an alternative to increasing this threshold, a post-processing procedure may efficiently eliminate most false positives by more accurately estimating the corresponding values of the measure.

\section{Variants and extensions}\label{sec:variants}

In this section we mention a number of ways in which our results can be extended. 

\subsection{Alternative filtering of false positives}
The threshold of $\mu/2$ in the {\sc BiSam} algorithms means that we filter away most pairs whose similarity is far from $\Delta$. An alternative is to spend more time on the pairs $(i,j)\in M$, using a sampling method to obtain a more accurate estimate of $|S_i\cap S_j|$.
A suitable technique could be to use min-wise independent hash functions~\cite{BroderCFM00, Indyk99} to obtain a sketch of each set $S_i$. 
It suffices to compare two sketches in order to have an approximation of the jaccard similarities of $S_i$ and $S_j$, which in turn gives an approximation of $|S_i \cap S_j|$. Based on this we may decide if a pair is likely to be interesting, or if it is possible to filter it out. The sketches could be built and maintained during the \textsc{ItemCount} procedure using, say, a logaritmic number of hash functions. Indyk~\cite{Indyk99} presents an efficient class of (almost) min-wise independent hash functions.

For some similarity measures such as {\em lift\/} and {\em overlap coefficient\/} the similarity of two sets may be high even if the sets have very different sizes. In such cases, it may be better to sample the smaller set, say, $S_i$, and use a hash table containing the larger set $S_j$ to estimate the fraction $|S_i\cap S_j|/|S_i|$.

\subsection{Reducing space usage by using counting Bloom filters}
At the cost of an extra pass over the data, we may reduce the space usage. The idea, previously found in e.g.~\cite{park.ea:effective-hash-based:95}, is to initially create an approximate representation of $M$ using counting Bloom filters (see~\cite{BM:02} for an introduction). Then, in a subsequent pass we may count only those pairs that, according to the approximation, may occur at least $\mu/2$ times.

\subsection{Weighted items}
Some applications of the cosine measure, e.g.~in information retrieval, require the items to be weighted. {\sc BiSam} easily extends to this setting.

\subsection{Adaptive variant.}

Instead of letting $\Delta$ be a user-defined variable, we may (informally) let $\Delta$ go from $\infty$ towards~$0$. This can be achieved by maintaining a priority queue of item pairs, where the priority reflects the value of $\Delta$ that would allow the pair to be sampled. Because $f$ is non-increasing in all parameters it suffices to have a linear number of pairs from each transaction in the priority queue at any time, namely the pairs that are ``next in line'' to be sampled. For each of the similarity measures in Fig.~\ref{table:functions} the value of $\Delta$ for a pair $(i,j)$ is easily computed by solving the equation $f(|S_i|,|S_j|,s) \mu=r$ for $s$. Decreasing $\Delta$ corresponds to removing the pair with the maximum value from the priority queue. At any time, the set of sampled item pairs will correspond exactly to the choice of $\Delta$ given by the last pair extracted from the priority queue. The procedure can be stopped once sufficiently many results have been found.

\subsection{Composite measures}

Notice that if $f_1(|S_i|,|S_j|,\Delta)$ and $f_2(|S_i|,|S_j|,\Delta)$ are both non-increasing, then any linear combination $\alpha f_1 + \beta f_2$, where $\alpha, \beta > 0$, is also non-increasing. Similarly, $\min(\alpha f_1,\beta f_2)$ is non-increasing. This allows us to use {\sc BiSam} to directly search for pairs with high similarity according to several measures (corresponding to $f_1$ and $f_2$), e.g., pairs with cosine similarity at least $0.7$ {\em and} lift at least~$2$.

\section{Experiments}

To make experiments fully reproducible and independent of implementation details and machine architecture, we focus our attention on the number of hash table operations, and the number of items in the hash tables. That is, the time for {\sc BiSam} is the number of items in the input set plus the number of pairs inserted in the multiset $M$. The space of {\sc BiSam} is the number of distinct items (for support counts) plus the number of distinct pairs in $M$. Similarly, the time for methods based on exact counting is the number of items in the input set plus the number of pairs in all transactions (since every pair is counted), and the space for exact counting is the number of distinct items plus the number of distinct pairs that occur in some transaction.

We believe that these simplified measures of time and space are a good choice for two reasons. First, hash table lookups and updates require hundreds of clock cycles unless the relevant key is in cache. This means that a large fraction of the time spent by a well-tuned implementation is used for hash table lookups and updates. Second, we are comparing two approaches that have a similar behavior in that they count supports of items and pairs. The key difference thus lies in the number of hash table operations, and the space used for hash tables. Also, this means that essentially any speedup or space reduction applicable to one approach is applicable to the other (e.g.~using counting Bloom filters to reduce space usage).

\subsection{Data sets}

Experiments have been run on both real datasets and artificial ones. We have used most of the datasets of the Frequent Itemset Mining Implementations (FIMI) Repository\footnote{\texttt{http://fimi.cs.helsinki.fi/}}. In addition, we have created three data sets based on the internet movie database (IMDB).

Fig.~\ref{table:data-description} contains some key figures on the data sets.

\begin{figure*}[!t]
\begin{center}
\begin{tabular}{c||c|c|c|c|c|}
\raisebox{-1.5ex}[0pt][0pt]{\bf Dataset} & {\bf distinct} & {\bf number of} & {\bf avg.~trans-} & {\bf max.~trans-} & {\bf avg.~items}\\
 	 & {\bf items} & {\bf transactions} & {\bf action size} & {\bf action size} & {\bf support}\\
\hline
\hline
Chess & 75 & 3196 & 37 & 37 & 1577\\
Connect & 129 & 67555 & 43 & 43 & 22519\\
Mushroom & 119 & 8134 & 23 & 23 & 1570\\
Pumsb & 2113 & 49046 & 74 & 74 & 1718\\
Pumsb\_star & 2088 & 49046 & 50 & 63 & 1186\\
Kosarak & 41270 & 990002 & 8 & 2498 & 194\\
BMS-WebView-1 & 497 & 5962 & 2 & 161 & 301\\
BMS-WebView-2 & 3340 & 59602 & 2 & 161 & 107\\
BMS-POS & 1657 & 515597 & 6 & 164 & 2032\\
Retail & 16470 & 88162 & 10 & 76 & 55\\
Accidents & 468 & 340183 & 33 & 51 & 24575\\
T10I4D100K & 870 & 100000 & 10 & 29 & 1161\\
T40I10D100K & 942 & 100000 & 40 & 77 & 4204\\
\hline
actors & 128203 & 51226 & 31 & 1002 & 12\\
directorsActor & 51226 & 3783 & 1221 & 8887 & 90\\
movieActors & 50645 & 133633 & 12 & 2253 & 33\\
\hline
\end{tabular}
\end{center}
\caption{\normalsize\em Key figures on the data sets used for experiments. The first 13 data sets are from the FIMI repository. The last 3 were extracted from the May 29, 2009 snapshot of the Internet Movie Database (IMDB).
The datasets {\bf Chess}, {\bf Connect}, {\bf Mushroom}, {\bf Pumsb}, and {\bf Pumsb\_star} were prepared by Roberto Bayardo from the UCI datasets and PUMBS. {\bf Kosarak} contains (anonymized) click-stream data of a hungarian on-line news portal, provided by Ferenc Bodon. {\bf BMS-WebView-1}, {\bf BMS-WebView-2}, and {\bf BMS-POS} contain clickstream and purchase data of a legwear and legcare web retailer, see~\cite{kddcup2000} for details.
{\bf Retail} contains the (anonymized) retail market basket data from a Belgian retail store~\cite{KDD99*254}. {\bf Accidents} contains (anonymized) traffic accident data~\cite{geurts03using}. The datasets {\bf T10I4D100K} and {\bf T10I4D100K} have been generated using an IBM generator from the Almaden Quest research group. {\bf Actors} contains the set of rated movies for each male actor who has acted in at least 10 rated movies. {\bf DirectorActor} contains, for each director who has directed at least 10 rated movies, the set of actors from {\bf Actors} that this director has worked with in rated movies. {\bf MovieActor} is the inverse relation of {\bf Actors}, listing for each movie a set of actors.}\label{table:data-description}
\end{figure*}

%The IMDB data files have been created as follows. We downloaded the files with the following headers:
%
%CRC: 0x5E7DB98B  File: ratings.list  Date: Fri May 29 01:00:00 2009
%CRC: 0x623B75F4  File: actors.list  Date: Fri May 29 01:00:00 2009
%CRC: 0xF21F74A9  File: directors.list  Date: Fri May 29 01:00:00 2009
%
%imdb2fimi.java:
%- We create a hash table containing the names and years of movies in ratings.list, and associate a unique number with each.
%- Then all actors in actors.list who have at least 10 movies in ratings are selected for inclusion in actors.dat. Each line lists the numbers of the movies (only those in ratings.list) for an actor.
%- In exactly the same way, we make a file directors.dat with movies for each director who has directed at least 10 movies.
%
%CreateDirectorActor.java:
%- Put the data from actors.dat and directors.dat into MySQL (assign IDs to each)
%- Run a join that computes the director-actor relationship
%- Extract a file, DirectorActor.dat, where each line is the IDs of actors that a given director has worked with.

\subsection{Results and discussion}

Fig.~\ref{table:finaltable} shows the results of our experiments for the cosine measure. The time and space for {\sc BiSam} is a random variable. The reported number is an {\em exact\/} computation of the expectation of this random variable. Separate experiments have confirmed that observed time and space is relatively well concentrated around this value. The value of $\Delta$ used is also shown --- it was chosen manually in each case to give a ``human readable'' output of around 1000 pairs.  (For the IMDB data sets and the Kosarak data set this was not possible; for the latter this behaviour was due to a large number of false positives.)
Note that choosing a smaller~$\Delta$ would bring the performance of {\sc BiSam} closer to the exact algorithms; this is not surprising, since lowering $\Delta$ means reporting pairs having a smaller similarity measure, increasing in this way the number of samples taken. As noted before, we are usually interested in reporting pairs with high similarity, for almost any reasonable scenario.

The results for the other measures are omitted for space reasons, since they are very similar to the ones reported here. This is because the complexity of {\sc BiSam} is, in most cases, dominated by the first phase (counting item frequencies), meaning that fluctuations in the cost of the second phase have little effect. This also suggests that we could increase the value of $\mu$ (and possibly increase the value of the threshold $\mu/2$ used in the {\sc BiSam} algorithm) without significantly changing the time complexity of the algorithm.

\begin{figure*}[!t]
 \begin{center}
   \subfigure[\normalsize\em Comparison of the time for {\sc BiSam} and for exact counting in all experiments. The line is the identity function. Typical difference is about an order of magnitude.]
   {\label{fig:time-comparison}\includegraphics[width=.47\linewidth]{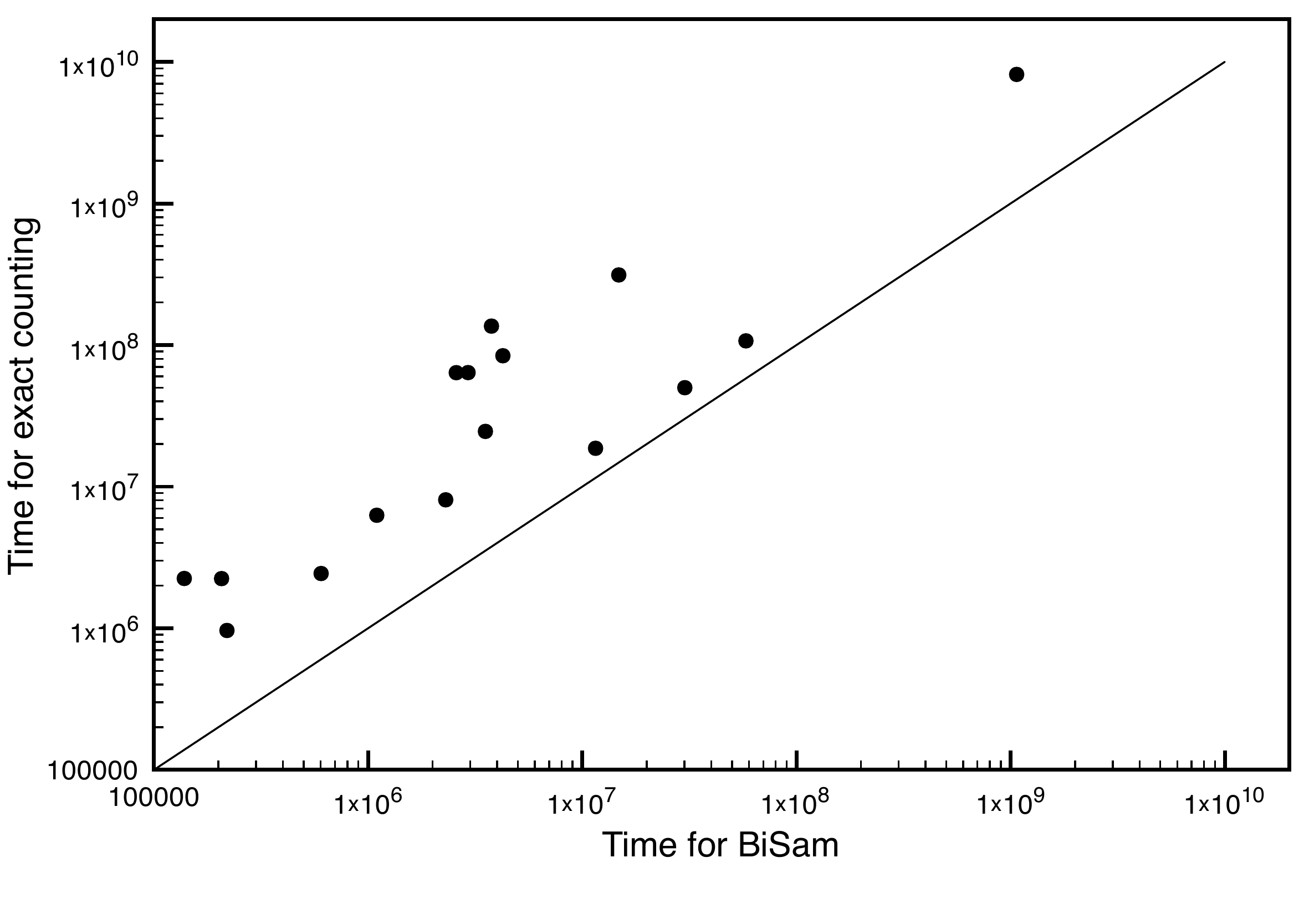}}\qquad\;
   \subfigure[\normalsize\em Comparison of the space for {\sc BiSam} and for exact counting in all experiments. The line is the identity function.]{\label{fig:space-comparison}\includegraphics[width=.47\linewidth]{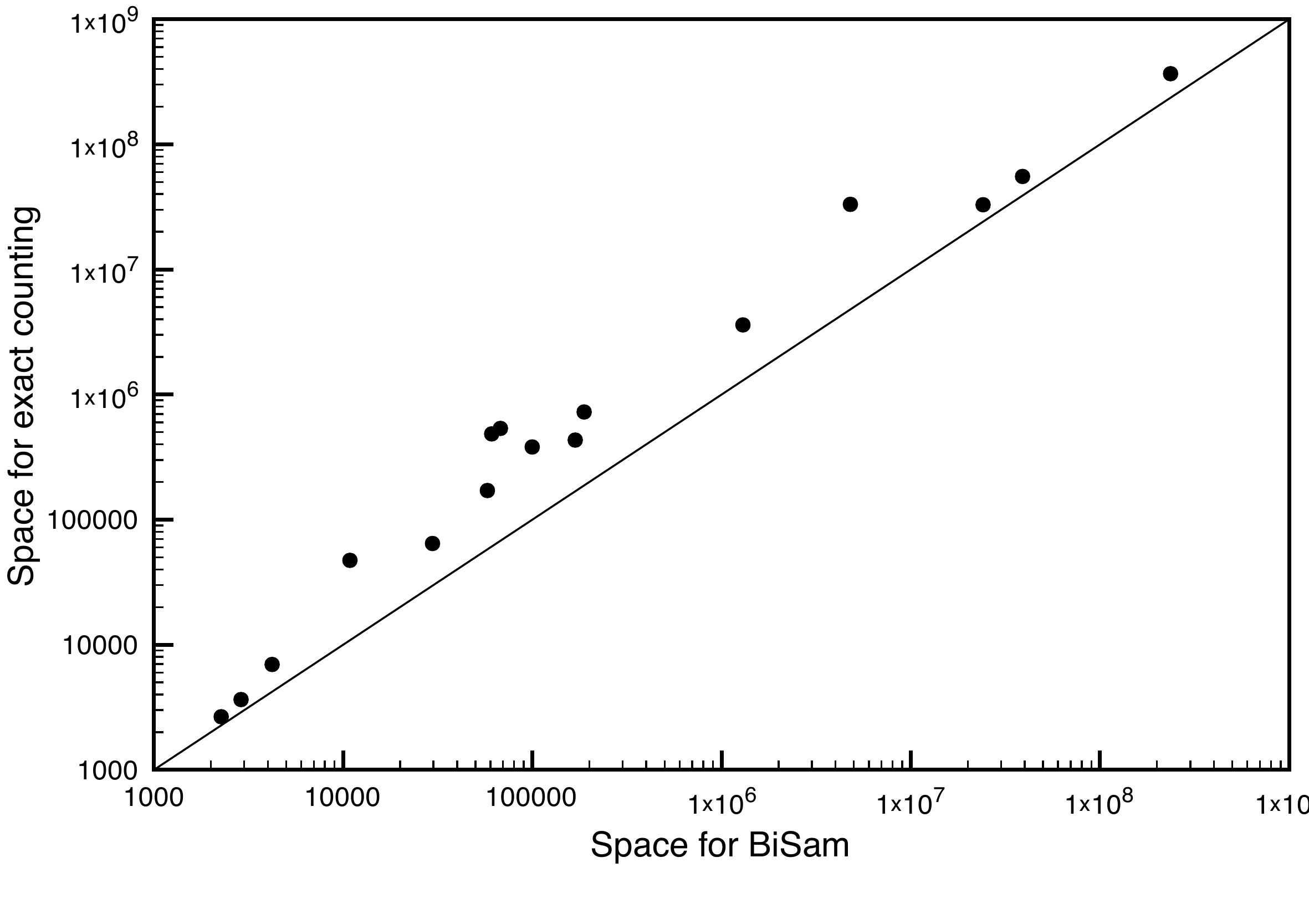}}
\caption{\normalsize\em Space and time comparisons}
\end{center}
\end{figure*}

We see that the speedup obtained in the experiments varies between a factor 2 and a factor over 30. Figures~\ref{fig:time-comparison} and \ref{fig:space-comparison} give a graphical overview. The largest speedups tend to come for data sets with the largest average transaction size, or data sets where some transactions are very large (e.g.~{\bf Kosarak}). However, as our theoretical analysis suggests, large transaction size alone is not sufficient to ensure a large speedup --- items also need to have support that is not too small. So while the {\bf DirectorActor} data set has very large average transaction size, the speedup is only moderate because the support of items is low. In a nutshell, {\sc BiSam} gives the largest speedups when there is a combination of relatively large transactions and relatively high average support. The space usage of {\sc BiSam} ranges from being quite close to the space usage for exact counting, to a decent reduction.

Though we have not experimented with methods based on locality-sensitive hashing (LSH), we observe that our method appears to have an advantage when the number $n$ of distinct items is large. This is because LSH in general (and in particular for cosine similarity) requires comparison of $\binom{n}{2}$ pairs of hash signatures. For the data sets {\bf Retail}, {\bf BMS-Webview-2}, {\bf Actors}, and {\bf MovieActors} the ratio between the number of signature comparisons and the number of hash table operations required for {\sc BiSam} is in the range 9--265. While these numbers are not necessarily directly comparable, it does indicate that {\sc BiSam} has the potential to improve LSH-based methods that require comparison of all signature pairs.

% Actors: n=128203, time 31000000, factor 265
% MovieActors: n=50645, time 58100000, factor 22
% Retail: n=16470, time 23 * 10^5, factor 59
% BMS WebView 2: n=3340, time 6 * 10^5, factor 9.3
% Kosarak: n=41280, n^2=1704038400, time 148*10^5, factor 5.75

\begin{figure*}[!t]
\begin{center}
\begin{tabular}{c||r|r|r|r|r|r|c|c|}
\hline
	 & \multicolumn{3}{|c|}{{\bf Time}} & \multicolumn{3}{|c|}{{\bf Space}} & \multicolumn{2}{|c|}{} \\
	 \hline
	 {\bf Dataset} & {\sc BiSam\phantom{..}} & Exact counting & Ratio & {\sc BiSam\phantom{a}} & Exact counting & Ratio & $\Delta$ & \#output\\ 
	 \hline
	 \hline
Chess & $1.39\cdot 10^5$ & $22.5\cdot 10^5$ & 16.21 & $2.27\cdot10^3$ & $2.66\cdot10^3$ & 1.17 & 0.6 & 1039\\
Connect & $29.3\cdot 10^5$ & $639\cdot 10^5$ & 21.82 & $4.21\cdot10^3$ & $6.96\cdot10^3$ & 1.65 & 0.7 & 1025\\
Mushroom & $2.07\cdot 10^5$ & $22.4\cdot 10^5$ & 10.84 & $2.89\cdot10^3$ & $6.29\cdot10^3$ & 2.18 & 0.4 & 976\\
Pumsb & $37.7\cdot 10^5$ & $1360\cdot 10^5$ & 36.14 & $67.8\cdot10^3$ & $536\cdot10^3$ & 7.91 & 0.7 & 3070\\
Pumsb\_star & $25.8\cdot 10^5$ & $638\cdot 10^5$ & 24.74 & $60.9\cdot10^3$ & $485\cdot10^3$ & 7.95 & 0.7 & 1929\\
Kosarak & $148\cdot 10^5$ & $3130\cdot 10^5$ & 21.13 & $4790\cdot10^3$ & $33100\cdot10^3$ & 6.92 & 0.95 & 63500\\
BMS-WebView-1 & $2.19\cdot 10^5$ & $9.64\cdot 10^5$ & 4.40 & $29.7\cdot10^3$ & $64.5\cdot10^3$ & 2.17 & 0.4 & 1226\\
BMS-WebView-2 & $6.03\cdot 10^5$ & $24.4\cdot 10^5$ & 4.04 & $188\cdot10^3$ & $725\cdot10^3$ & 3.86 & 0.6 & 1317\\
BMS-POS & $35.3\cdot 10^5$ & $246\cdot 10^5$ & 6.96 & $99.8\cdot10^3$ & $381\cdot10^3$ & 3.82 & 0.15 & 1263\\
Retail & $23\cdot 10^5$ & $80.7\cdot 10^5$ & 3.50 & $1300\cdot10^3$ & $3600\cdot10^3$ & 2.78 & 0.3 & 1099\\
Accidents & $115\cdot 10^5$ & $187\cdot 10^5$ & 1.62 & $10.9\cdot10^3$ & $47.3\cdot10^3$ & 4.35 & 0.5 & 995\\
T10I4D100K & $11\cdot 10^5$ & $62.8\cdot 10^5$ & 5.72 & $57.8\cdot10^3$ & $171\cdot10^3$ & 2.95 & 0.4 & 846\\
T40I10D100K & $42.6\cdot 10^5$ & $841\cdot 10^5$ & 19.74 & $168\cdot10^3$ & $433\cdot10^3$ & 2.57 & 0.5 & 1120\\
\hline
Actors & $301\cdot 10^5$ & $500\cdot 10^5$ & 1.66 & $24100\cdot10^3$ & $32900\cdot10^3$ & 1.37 & 0.5 & 18531\\
DirectorsActor & $10700\cdot 10^5$ & $81500\cdot 10^5$ & 7.64 & $236000\cdot10^3$ & $367000\cdot10^3$ & 1.56 & 0.5 & ---\\
MovieActors & $581\cdot 10^5$ & $1070\cdot 10^5$ & 1.84 & $38900\cdot10^3$ & $55400\cdot10^3$ & 1.42 & 0.5 & 43567\\
\hline
\end{tabular}
\end{center}
\caption{\normalsize\em Result of experiments for the cosine measure and $\mu=15$ (which gives false negative probability 1.8\%). \#output is the number of pairs of items reported; $\Delta$ is the threshold for ``interesting similarity.'' DirectorsActor lacks the output because of the huge number of pairs.}\label{table:finaltable}
\end{figure*}

\section{Conclusion}

We have presented a new sampling-based method for finding associations in data. Besides our initial experiments, indicating that large speedups may be obtained, there appear to be many opportunities for using our approach to implement association mining systems with very high performance. Some such opportunities are outlined in Section~\ref{sec:variants}, but many nontrivial aspects would have to be considered to do this in the best way.

\bigskip

\section*{Acknowledgment}
We wish to thank Blue Martini Software for contributing the KDD Cup 2000 data. Also, we thank the reviewers of the ICDM submission for pointing out several related works.

\IEEEtriggeratref{24}


\bibliographystyle{IEEEtran}

\bibliography{IEEEabrv,mining}

\end{document}